\documentclass{pasj00}

\begin{document}
\SetRunningHead{T. Kato}{Detection of Short Fading Episodes in Two Dwarf Novae}

\Received{}
\Accepted{}

\title{Detection of Short Fading Episodes in Two Dwarf Novae from VSNET Observations}

\author{Taichi \textsc{Kato}}
\affil{Department of Astronomy, Kyoto University,
       Sakyo-ku, Kyoto 606-8502}
\email{tkato@kusastro.kyoto-u.ac.jp}


\maketitle

\begin{abstract}
   From the observations reported to VSNET, we detected short fading
episodes in two dwarf novae, RX And (orbital period 5.0 hr) and
SU UMa (orbital period 1.8 hr).  The episodes in RX And can be naturally
explained as the response to a short-term reduction of the mass-transfer
rate.  A qualitative comparison with VY Scl-type fadings is also
discussed.  In SU UMa, the same explanation is expected to be more
difficult to apply.  The viable, but still poorly understood,
possibilities include the temporary reduction of quiescent disk viscosity
and the temporary reduction of mass-transfer rate.  If the latter
possibility is confirmed, we probably need a different mechanism from
that of VY Scl-type stars.
\end{abstract}

\section{Introduction}

   Cataclysmic variables (CVs) are close binary systems consisting of
a white dwarf and a red-dwarf secondary transferring matter via the Roche
lobe overflow.  Among CVs, there exist a group of objects called
VY Scl-type stars or anti-dwarf novae (\cite{war95book}).
In VY Scl-type stars, the mass-transfer from the secondary is occasionally
reduced, or possible even stops \citep{rob81mvlyr}.  Well-described
examples of typical VY Scl-type stars can be found in
\cite{rob81mvlyr}; \cite{wen83mvlyr}; \cite{fuh85mvlyr}; \cite{gre98vyscl};
\cite{kat02kraur}.  Fading episodes (low states) similar to those of
VY Scl-type stars are known to appear in various kinds of CVs:
intermediate polars \citep{gar88vyscldqher};
some dwarf novae (\cite{sio01rxand}; \cite{kat02rxand}; \cite{sch02rxand}).
Polars and supersoft X-ray source are also known to show low states
\citep{hes00CVlowstate}, but they are likely to have different origin
(for supersoft X-ray source, see e.g. \cite{hac03SSS}).  Past studies
of temporary reduction of mass-transfer rates in CVs mostly focused on
episodes with long time scales (usually $\geq$100 d).  We here report
the detection of much shorter (tens of days) fading episodes, at least
some of which are likely arising from temporary reduction of mass-transfer
rates, in two dwarf novae RX And and SU UMa.
The analyzed data are from visual observations reported to
VSNET\footnote{
$\langle$http://www.kusastro.kyoto-u.ac.jp/vsnet/$\rangle$.
}

\section{RX Andromedae}\label{sec:rx}

   RX And is one of the prototypical Z Cam-type dwarf novae which show
standstills in addition to ordinary dwarf nova outbursts
see e.g. \cite{hel01book} Sect. 5.4; see also \cite{war74vysclzcam};
\cite{mey83zcam}; \cite{opp98zcam}).
Most recently, RX And has been shown to be an unique Z Cam-type dwarf
nova which also exhibits occasional ``low states'' (\cite{sio01rxand};
\cite{kat02rxand}; \cite{sch02rxand}).  \citet{sch02rxand} even
suggested that RX And is a transitional object between Z Cam-type dwarf
novae and VY Scl-type stars (cf. \cite{war74vysclzcam}; \cite{war95book};
\cite{gre98vyscl}), although \citet{kat02rxand} reported difference
between fadings in RX And and in VY Scl-type stars.

   \citet{sch02rxand} examined the historical light curve of RX And
and reported that typical fadings of RX And lasted for
$\sim$100 d.\footnote{
Before the discovery of the 1996 fading episode, RX And in deep quiescence
(or in low state) may have been confused with the nearby star.  Because of
the ambiguity in interpreting the historical data, both \citet{kat02rxand}
and apparently \citet{sch02rxand} ascribed ``intervals without
outbursts'' to historical fading episodes.}

\begin{figure*}
  \begin{center}
    \FigureFile(160mm,100mm){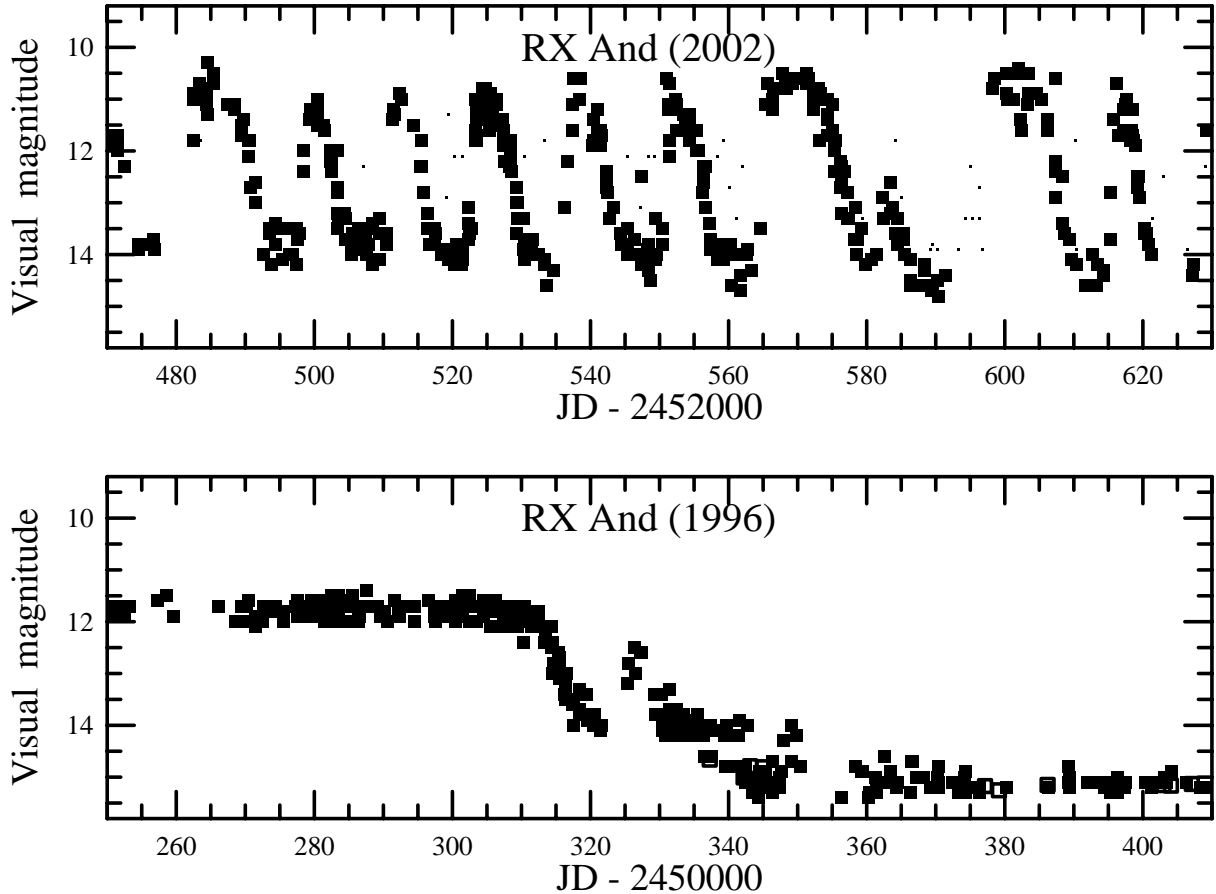}
  \end{center}
  \caption{(Upper:) Light curve of RX And in 2002.
  While the object initially showed regular dwarf nova outbursts,
  it entered a short faint state starting at around JD 2452570.
  (Lower:) Long fading episode in 1996.  The object entered a faint state
  at around JD 2450310, after the initial state of a standstill.
  An occurrence of a short, faint brightening just following the start
  of the fading episode is remarkably similar on both occasions, although
  the duration of the fading is very different.  The large and small dots
  represent positive and upper limit visual observations reported to VSNET.
  The open squares represent CCD $V$-band observations from
  \citet{kat02rxand}.
  }
  \label{fig:rx}
\end{figure*}

   Here we report on the detection of an unusually short fading episode
in 2002.  The upper panel of figure \ref{fig:rx} shows the light curve
covering this event starting at around JD 2452570.
The errors of visual observations are usually less than 0.3 mag, which
do not affect the discussion.  About 3--4 d after reaching a
temporary minimum, the object underwent a short, small brightening
(JD 2452582).  This behavior is remarkably similar to a small brightening
when RX And entered a long faint state in 1996
(figure \ref{fig:rx}, lower panel).
Such behavior is just what is expected for a thermally stable accretion
disk suffering from a sudden decrease of the mass-transfer rate
(\cite{hon94v794aql}; \cite{kin98DI}).  We therefore identify
the 2002 phenomenon as an extremely short fading episode.  The recovery
from this fading took less than 30 d (figure \ref{fig:rx}, upper panel).

   No similarly short fading episode was identified in \citet{sch02rxand},
although there was a slight indication of a short fading around
JD 2449300.  No other similar event has been recorded since the firm
establishment in 1996 of the fading phenomenon in RX And.  From these
findings, such a short fading episode is extremely rare.

\section{SU Ursae Majoris}\label{sec:su}

\begin{figure*}
  \begin{center}
    \FigureFile(160mm,220mm){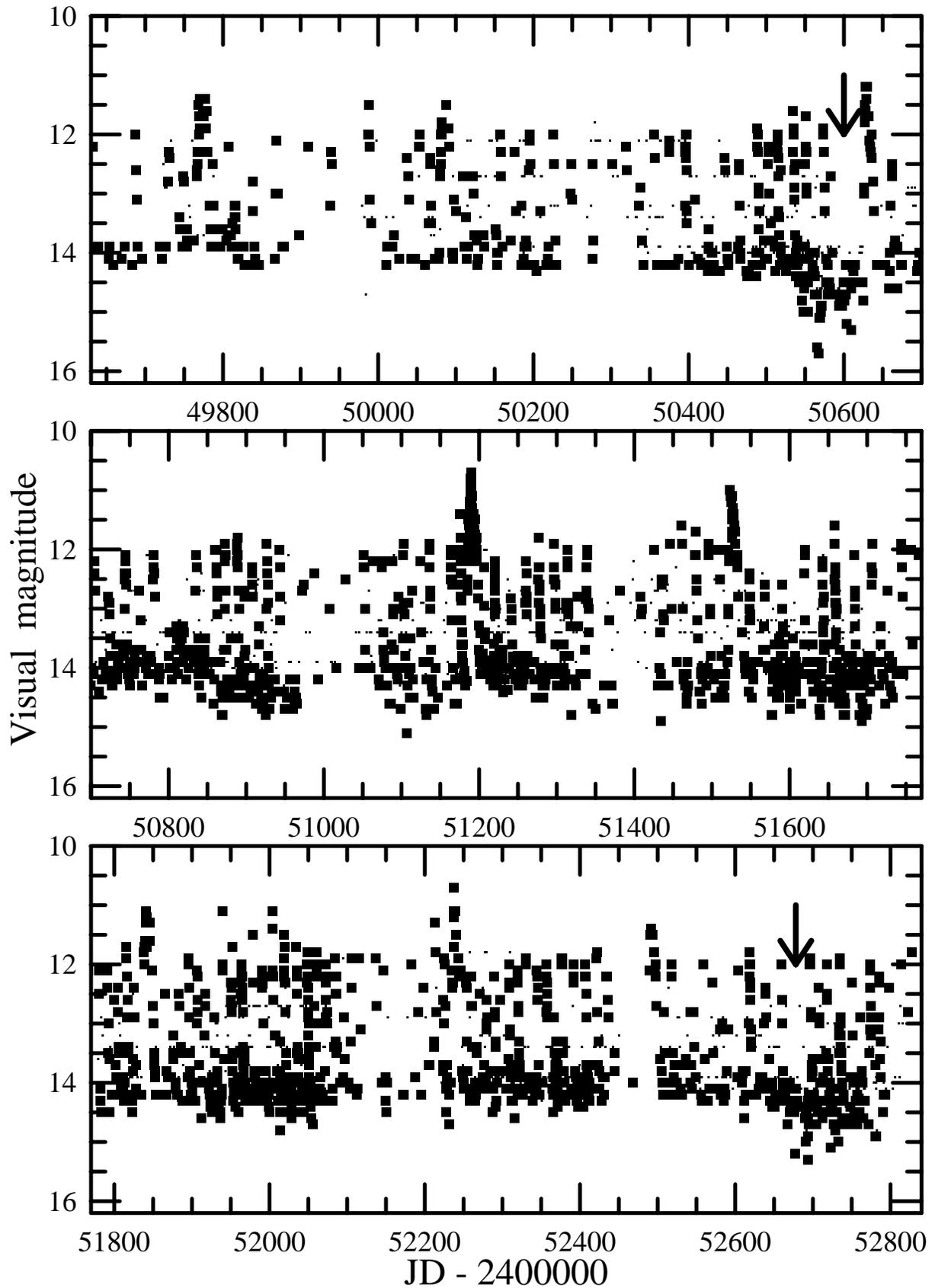}
  \end{center}
  \caption{Long-term light curve of SU UMa from VSNET observations.
  The large and small filled squares represent positive and negative
  (upper limit) observations.
  Two short fading episodes are shown with the thick arrows.
  }
  \label{fig:su}
\end{figure*}

\begin{figure}
  \begin{center}
    \FigureFile(88mm,60mm){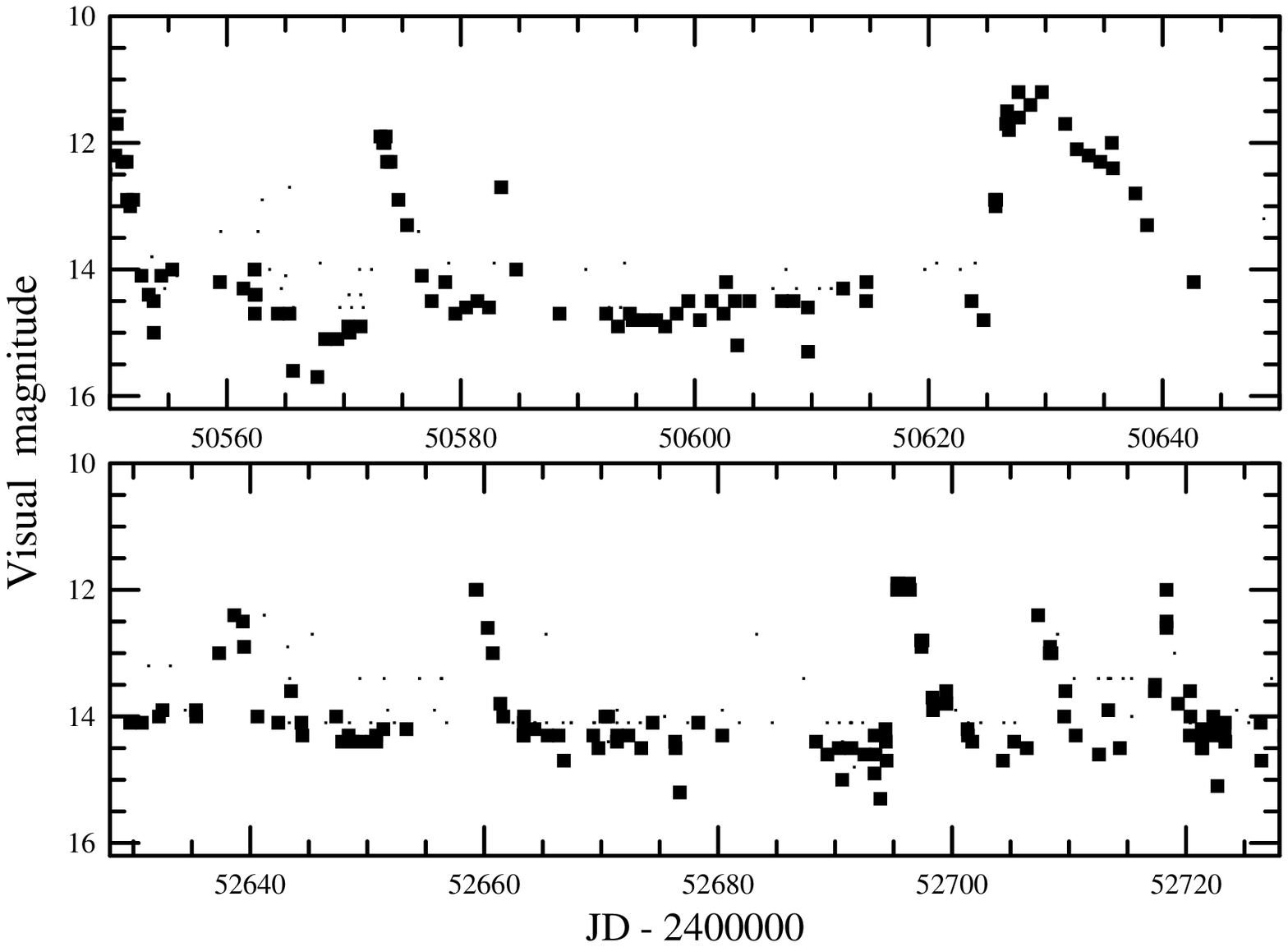}
  \end{center}
  \caption{Enlargements of two episodes of SU UMa.  The symbols are
  as in figure \ref{fig:su}.
  }
  \label{fig:sularge}
\end{figure}

   SU UMa is the well-known prototype of the SU UMa-type dwarf novae
\citep{uda90suuma}.  Although some irregularities in the long-term
light variation has been reported (e.g. \cite{ros00suuma};
\cite{kat02suuma}), past studies mainly pointed out the irregularities
in the supercycle (or occurrence of superoutbursts) and the occasional
absence of normal outbursts.  From the VSNET observation, we report on
the detection of short fadings, which may by analogous to that of
RX And (section \ref{sec:rx}).  Figure \ref{fig:su} shows the long-term
(1994--2003) light curve of SU UMa.  Although the object spends most
of its quiescence at around $V$=14.0, there are at least two
instances [marked with the arrows: around JD 2450600 (1997 May) and
JD 2452680 (2003 February)] when the object was transiently
observed below $V\sim$15.0.  During these phases, the number of outbursts
was remarkably reduced (figure \ref{fig:sularge}).
After the 1997 May fading, the object underwent a superoutburst.

\section{Discussion}

  In most VY Scl-type stars, the fading episodes (low states) usually
take much longer durations (cf. \cite{gre98vyscl}; \cite{kat02kraur}).
This difference may be a result of a stronger heating effect on
the accretion disk in VY Scl-type stars \citep{lea99vyscl}.

   Following the calculation by \citet{lea99vyscl},
very short events of the reduced mass-transfer in VY Scl-type stars,
if any, may not appear as prominent fadings since a heating can thermally
stabilize the accretion disk.  \citet{lea99vyscl} reported that a white
dwarf temperature of 4$\times$10$^4$ K produces ionized inner disk,
thereby preventing dwarf nova-type disk instability from occurring at low
mass-transfer rates, and preventing the system from immediately reaching
a deep, low state.  We can expect that
the effect would be more obvious when the decrease of mass-transfer occurs
more slowly, and when the disk surface density is close to the thermal
stability.  Observational evidence for the suppression of the disk
instability in a VY Scl star, V425 Cas, can be found in
\citet{kat01v425cas}.
The presence of a short fading episode in RX And thus might be
an additional manifestation of the difference between the faint states
in RX And and VY Scl-type stars \citep{kat02rxand}.

   We must note that there have been, however, arguments that a heating
effect rather produces small ``echo outbursts'', which are different
from the conclusion by \citet{lea99vyscl} (\cite{ham00DNirradiation};
\cite{sch01ADpostnova}).  Following these interpretations,
the small brightening (JD 2452582) may be alternatively interpreted as
a variety of echo outbursts.
Since the physical reason of the difference in simulations has not been
fully clarified, we should leave it as an open question.

   We must note that a similar short fading was once observed in the
intermediate polar AO Psc \citep{gar88vyscldqher}, which has been
sometimes considered as a candidate VY Scl-type star.  The truncation of
the accretion disk by the strong magnetic field of an intermediate polar
may have enabled such a short fading episode observable in AO Psc.

   The phenomenon in RX And seems to be more easily understandable
with a temporary reduction in mass-transfer rate, since RX And has
already shown a much longer episode of reduced mass-transfer
(\cite{sio01rxand}; \cite{kat02rxand}; \cite{sch02rxand}).  Although
the exact cause of reduced mass-transfer in VY Scl-type stars is not
yet clarified, the starspot hypothesis \citep{liv94CVstarspot} would
be a promising explanation.  Since the starspot activity is expected
to be stronger in stars with higher magnetic activities, this explanation
would naturally explain the high occurrence of VY Scl-type stars
in only in systems of long orbital periods ($P_{\rm orb}$), particularly
in the range 3 hr$\leq P_{\rm orb}\leq$ 4hr (see \cite{liv94CVstarspot}
for the full explanation).  The location of RX And ($P_{\rm orb}$ = 5.0 hr)
is just above this region, and the proposed mechanism would reasonably
work.  The long-period dwarf novae with occasional low states include
WW Cet ($P_{\rm orb}$ = 4.2 hr, \cite{rin96wwcet}), although it is not
clear whether these low states actually correspond to reduced mass-transfer
phenomena.

   This explanation, however, is more difficult to apply to SU UMa
(section \ref{sec:su}).
Since SU UMa is an object with $P_{\rm orb}$ = 1.8 hr, the secondary
star is expected to become fully convective and we can not
{\it canonically} expect a high magnetic activity
(\cite{ver81magneticbraking};
\cite{rap82CVevolution}; \cite{rap83CVevolution}; \cite{ver84periodgap};
\cite{rap84ultrashortPbinary}; \cite{rit85CVperiodgap}).
The first fading episode (1997 May) might have a different origin. 
This episode was followed by a superoutburst, suggesting that the large
amount of mass must have stored \citep{osa89suuma} even during
this fading.
During this fading, the low quiescent brightness may have been a
temporary reduction of disk viscosity which also suppressed normal
outbursts \citep{kat02suuma}, but once the thermal instability occurs,
this would lead to a larger outburst (the extreme effect would be
best seen in WZ Sge-type stars: \cite{osa95wzsge},
\cite{osa03DNoutburst}).  The second fading (2003 February) was not
associated with similar occurrence of a superoutburst.  This phenomenon
may have been an episode with a true reduction of mass-transfer.
If this possibility of mass-transfer reduction is confirmed,
we probably need a different mechanism from that of VY Scl-type
stars.

    We must also note that some objects with $P_{\rm orb}$ similar to
SU UMa are claimed to have ``low states" (e.g.
HT Cas: $P_{\rm orb}$ = 1.8 hr, \cite{zha86htcas};
\cite{woo95htcasXray}; \cite{rob96htcas} and
IR Com: $P_{\rm orb}$ = 2.1 hr, \cite{ric95ircom}; \cite{kat02ircom}).
There are also SU UMa-type dwarf novae with long-term variation of
outburst properties (V1159 Ori: \cite{kat01v1159ori}; V503 Cyg:
\cite{kat02v503cyg}; DM Lyr: \cite{nog03dmlyr};
MN Dra = Var73 Dra: \cite{nog03var73dra}).
It is not still certain whether or not variable mass-transfer rate
is responsible for these phenomena in these short-$P_{\rm orb}$
dwarf novae.

\vskip 3mm

The authors are grateful to many observers who reported vital
observations to VSNET.
This work is partly supported by a grant-in-aid (13640239, 15037205)
from the Japanese Ministry of Education, Culture, Sports,
Science and Technology.


\begin{thebibliography}{}

\bibitem[Fuhrmann(1985)]{fuh85mvlyr}
  Fuhrmann, B.\ 1985, Inf. Bull. Variable Stars, 2833

\bibitem[Garnavich, Szkody(1988)]{gar88vyscldqher}
  Garnavich, P., \& Szkody, P.\ 1988, \pasp, 100, 1522

\bibitem[Greiner(1998)]{gre98vyscl}
  Greiner, J.\ 1998, \aap, 336, 626

\bibitem[Hachisu, Kato(2003)]{hac03SSS}
  Hachisu, I., \& Kato, M.\ 2003, \apj, in press, (astro-ph/0301489)

\bibitem[Hameury et~al.(2000)]{ham00DNirradiation}
  Hameury, J.-M., Lasota, J.-P., \& Warner, B.\ 2000, \aap, 353, 244

\bibitem[Hellier(2001)]{hel01book}
  Hellier, C.\ 2001, Cataclysmic Variable Stars: how and why they vary.
  (Berlin: Springer-Verlag)

\bibitem[Hessman(2000)]{hes00CVlowstate}
  Hessman, F.~V.\ 2000, New Astron. Rev., 44, 155

\bibitem[Honeycutt et~al.(1994)]{hon94v794aql}
  Honeycutt, R.~K., Cannizzo, J.~K., \& Robertson, J.~W.\ 1994, \apj, 425, 835

\bibitem[Kato(2001)]{kat01v1159ori}
  Kato, T.\ 2001, \pasj, 53, L17

\bibitem[Kato(2002)]{kat02suuma}
  Kato, T.\ 2002, \aap, 384, 206

\bibitem[Kato et~al.(2002a)]{kat02ircom}
  Kato, T., Baba, H., \& Nogami, D.\ 2002a, \pasj, 54, 79

\bibitem[Kato et~al.(2002b)]{kat02v503cyg}
  Kato, T., Ishioka, R., \& Uemura, M.\ 2002b, \pasj, 54, 1029

\bibitem[Kato et~al.(2002c)]{kat02kraur}
  Kato, T., Ishioka, R., \& Uemura, M.\ 2002c, \pasj, 54, 1033

\bibitem[Kato et~al.(2002d)]{kat02rxand}
  Kato, T., Nogami, D., \& Masuda, S.\ 2002d, \pasj, 54, 575

\bibitem[Kato et~al.(2001)]{kat01v425cas}
  Kato, T., Uemura, M., Ishioka, R., \& Kinnunen, T.\ 2001, \pasj, 53, 1185

\bibitem[King, Cannizzo(1998)]{kin98DI}
  King, A.~R., \& Cannizzo, J.~K.\ 1998, \apj, 499, 348

\bibitem[Leach et~al.(1999)]{lea99vyscl}
  Leach, R., Hessman, F.~V., King, A.~R., Stehle, R., \& Mattei, J.\ 1999,
  \mnras, 305, 225

\bibitem[Livio, Pringle(1994)]{liv94CVstarspot}
  Livio, M., \& Pringle, J.~E.\ 1994, \apj, 427, 956

\bibitem[Meyer, Meyer-Hofmeister(1983)]{mey83zcam}
  Meyer, F., \& Meyer-Hofmeister, E.\ 1983, \aap, 121, 29

\bibitem[Nogami et~al.(2003a)]{nog03dmlyr}
  Nogami, D., Baba, H., Matsumoto, K., \& Kato, T.\ 2003a, \pasj, 55, 483

\bibitem[Nogami et~al.(2003b)]{nog03var73dra}
  Nogami, D. {et~al.}\ 2003b, \aap, 404, 1067

\bibitem[Oppenheimer et~al.(1998)]{opp98zcam}
  Oppenheimer, B.~D., Kenyon, S.~J., \& Mattei, J.~A.\ 1998, \aj, 115, 1175

\bibitem[Osaki(1989)]{osa89suuma}
  Osaki, Y.\ 1989, \pasj, 41, 1005

\bibitem[Osaki(1995)]{osa95wzsge}
  Osaki, Y.\ 1995, \pasj, 47, 47

\bibitem[Osaki, Meyer(2003)]{osa03DNoutburst}
  Osaki, Y., \& Meyer, F.\ 2003, \aap, 401, 325

\bibitem[Rappaport, Joss(1984)]{rap84ultrashortPbinary}
  Rappaport, S., \& Joss, P.~C.\ 1984, \apj, 283, 232

\bibitem[Rappaport et~al.(1983)]{rap83CVevolution}
  Rappaport, S., Joss, P.~C., \& Verbunt, F.\ 1983, \apj, 275, 713

\bibitem[Rappaport et~al.(1982)]{rap82CVevolution}
  Rappaport, S., Joss, P.~C., \& Webbink, R.~F.\ 1982, \apj, 254, 616

\bibitem[Richter, Greiner(1995)]{ric95ircom}
  Richter, G.~A., \& Greiner, J.\ 1995, in Cataclysmic Variables, ed. A.
  Bianchini, M. Della~Valle \& M. Orio (Dordrecht: Kluwer Academic
  Publishers), ~177

\bibitem[Ringwald et~al.(1996)]{rin96wwcet}
  Ringwald, F.~A., Thorstensen, J.~R., Honeycutt, R.~K., \& Smith, R.~C.\ 1996,
  \aj, 111, 2077

\bibitem[Ritter(1985)]{rit85CVperiodgap}
  Ritter, H.\ 1985, \aap, 145, 227

\bibitem[Robertson, Honeycutt(1996)]{rob96htcas}
  Robertson, J.~W., \& Honeycutt, R.~K.\ 1996, \aj, 112, 2248

\bibitem[Robinson et~al.(1981)]{rob81mvlyr}
  Robinson, E.~L., Barker, E.~S., Cochran, A.~L., Cochran, W.~D., \& Nather,
  R.~E.\ 1981, \apj, 251, 611

\bibitem[Rosenzweig et~al.(2000)]{ros00suuma}
  Rosenzweig, P., Mattei, J., Kafka, S., Turner, G.~W., \& Honeycutt, R.~K.\
  2000, \pasp, 112, 632

\bibitem[Schreiber, G"{a}nsicke(2001)]{sch01ADpostnova}
  Schreiber, M.~R., \& G\"{a}nsicke, B.~T.\ 2001, \aap, 375, 937

\bibitem[Schreiber et~al.(2002)]{sch02rxand}
  Schreiber, M.~R., G\"{a}nsicke, B.~T., \& Mattei, J.~A.\ 2002, \aap,
  384, L6

\bibitem[Sion et~al.(2001)]{sio01rxand}
  Sion, E.~M., Szkody, P., Gaensicke, B., Cheng, F.~H., LaDous, C., \& Hassall,
  B.\ 2001, \apj, 555, 834

\bibitem[Udalski(1990)]{uda90suuma}
  Udalski, A.\ 1990, \aj, 100, 226

\bibitem[Verbunt(1984)]{ver84periodgap}
  Verbunt, F.\ 1984, \mnras, 209, 227

\bibitem[Verbunt, Zwaan(1981)]{ver81magneticbraking}
  Verbunt, F., \& Zwaan, C.\ 1981, \aap, 100, L7

\bibitem[Warner(1995)]{war95book}
  Warner, B.\ 1995, Cataclysmic Variable Stars (Cambridge: Cambridge
  University Press)

\bibitem[Warner, van Citters(1974)]{war74vysclzcam}
  Warner, B., \& van Citters, G.~W.\ 1974, Observatory, 94, 116

\bibitem[Wenzel, Fuhrmann(1983)]{wen83mvlyr}
  Wenzel, W., \& Fuhrmann, B.\ 1983, Mitteil. Ver\"{a}nderl. Sterne, 9, 175

\bibitem[Wood et~al.(1995)]{woo95htcasXray}
  Wood, J.~H., Naylor, T., Hassall, B. J.~M., \& Ramseyer, T.~F.\ 1995, \mnras,
  273, 772

\bibitem[Zhang et~al.(1986)]{zha86htcas}
  Zhang, E.-H., Robinson, E.~L., \& Nather, R.~E.\ 1986, \apj, 305, 740

\end{thebibliography}
\end{document}